\documentclass[manuscript]{aastex}
\usepackage{amsmath,natbib}
\usepackage{rotating}
\usepackage{emulateapj5}

\newcommand{\spider}{\emph{Spider}}
\newcommand{\blast}{\textsc{BLAST}}
\newcommand{\boomerang}{\textsc{BOOMERANG}}
\newcommand{\bicep}{\emph{Robinson/BICEP}} 

\slugcomment{submitted to ApJ}

\shorttitle{Spider Systematics}
\shortauthors{MacTavish et al.}

\begin{document}

\title{ Spider Optimization:  Probing the Systematics of a Large Scale B-Mode Experiment }

\author{C.~J.~MacTavish\altaffilmark{1}, P.~A.~R.~Ade\altaffilmark{2},E.~S.~Battistelli\altaffilmark{3}
S.~Benton\altaffilmark{8}, R.~Bihary\altaffilmark{3},
J.~J.~Bock\altaffilmark{5,6}, J.~R.~Bond\altaffilmark{1}, 
J.~Brevik\altaffilmark{5}, S.~Bryan\altaffilmark{4},
C.~R.~Contaldi\altaffilmark{9}, B.~P.~Crill\altaffilmark{1,8}, 
O.~Dor\'e\altaffilmark{1}, L.~Fissel\altaffilmark{8}, 
S.~R.~Golwala\altaffilmark{6}, M.~Halpern\altaffilmark{3}, 
G.~Hilton\altaffilmark{10}, W.~Holmes\altaffilmark{5}, 
V.~V.~Hristov\altaffilmark{6}, K.~Irwin\altaffilmark{10}, 
W.~C.~Jones\altaffilmark{5,6}, C.~L.~Kuo\altaffilmark{6}, 
A.~E.~Lange\altaffilmark{6}, C.~Lawrie\altaffilmark{4}, 
T.~G.~Martin\altaffilmark{11}, P.~Mason\altaffilmark{6}, 
T.~E.~Montroy\altaffilmark{4}, C.~B.~Netterfield\altaffilmark{7,8}, 
D.~Riley\altaffilmark{4}, J.~E.~Ruhl\altaffilmark{4}, 
A.~Trangsrud\altaffilmark{6}, C.~Tucker\altaffilmark{2}, 
A.~Turner\altaffilmark{5}, M.~Viero\altaffilmark{8}, and D.~Wiebe\altaffilmark{7}
}

\altaffiltext{1}{Canadian Institute for Theoretical Astrophysics (CITA),University of Toronto, ON, Canada}
\altaffiltext{2}{School of Physics and Astronomy, Cardiff University, Wales, UK}
\altaffiltext{3}{Department of Physics and Astronomy, University of British Columbia, Vancouver, BC, Canada}
\altaffiltext{4}{Department of Physics, Case Western Reserve University, Cleveland, OH, USA}
\altaffiltext{5}{Jet Propulsion Laboratory, Pasadena, CA, USA}
\altaffiltext{6}{Department of Physics, California Institute of Technology, Pasadena, CA, USA}
\altaffiltext{7}{Department of Physics, University of Toronto, ON, Canada}
\altaffiltext{8}{Department of Astronomy and Astrophysics, University of Toronto, ON,Canada}
\altaffiltext{9}{Theoretical Physics, Blackett Laboratory, Imperial College, London, UK}
\altaffiltext{10}{National Institute of Standards and Technology, Boulder, CO, USA}
\altaffiltext{11}{Department of Mechanical and Industrial Engineering, University of Toronto, ON,Canada}

\begin{abstract}
\spider\ is a long-duration, balloon-borne polarimeter designed to 
measure large scale Cosmic Microwave Background (CMB) polarization with
very high sensitivity and control of systematics.  The instrument will
map over half the sky with degree angular resolution in I, Q and U Stokes
parameters, in four frequency bands from $96$ to $275$ GHz.  \spider's
ultimate goal is to detect the primordial gravity wave signal imprinted
on the CMB B-mode polarization.  One of the challenges in achieving this
goal is the minimization of the contamination of B-modes by systematic
effects. This paper explores a number of instrument systematics and
observing strategies in order to optimize B-mode sensitivity.  This is
done by injecting realistic-amplitude, time-varying systematics in a set
of simulated time-streams. Tests of the impact of detector noise
characteristics, pointing jitter, payload pendulations, polarization
angle offsets, beam systematics and receiver gain drifts are shown.
\spider's default observing strategy is to spin continuously in 
azimuth, with polarization modulation achieved by either a rapidly 
spinning half-wave plate or a rapidly spinning gondola and a slowly 
stepped half-wave plate.  Although the latter is more susceptible to 
systematics, results shown here indicate that either mode of operation 
can be used by \spider.
\end{abstract}


\section{Introduction}\label{sec:intro}

In the past decade, a wealth of data have pointed to a ``standard
model'' of the Universe, composed of $\sim 5\%$ ordinary matter, $\sim
22\%$ dark matter and $\sim 73\%$ dark energy in a flat geometry.  The
flatness of the Universe, the near isotropy of the CMB, and the
nearly-scale-invariant nature of the primordial scalar perturbations
from which structure grew support the existence of an early accelerating
phase dubbed ``inflation''. A necessary by-product of inflation is
tensor perturbations from quantum fluctuations in gravity waves.  A
detection of this Cosmological Gravity-Wave Background (CGB) would
give strong evidence of an inflationary period and determine its
energy scale, while a powerful upper limit would point to more radical
inflationary scenarios, e.g., involving string theory, or some
non-inflationary explanation of the observations.

The CGB imprints a unique signal in the curl-like, or B-mode, component
of the polarization of the CMB; detection of a B-mode signal can be used
to infer the presence of a CGB at the time of decoupling.  Direct
detection of the gravity waves is many decades off; an advanced Big Bang
Observer successor to LISA has been suggested as a way to achieve this
\citep{BBO1,BBO2}. Thus a measurement of the primordial B-modes is the
only feasible near-term way to detect the CGB and have a new window to
the physics of the early Universe \citep{Bock:2006}.

A CGB with a potentially measurable amplitude is a by-product of the
simplest models of single field inflation which can reproduce the
scalar spectral tilt observed in current combined CMB
data~\citep{spergel2007, MacTavish:2005yk}. Examples are chaotic
inflation from power law inflaton
potentials~\citep{linde:1983,linde:2005} or natural inflation from
cosine inflaton potentials involving angular (axionic) degrees of
freedom~\citep{adams93}. The amplitude is usually parameterized in terms
of the ratio of the tensor power spectrum to the scalar power spectrum,
$r={\cal P}_t (k_p)/{\cal P}_s (k_p)$, evaluated at a comoving
wavenumber pivot $k_p$, typically taken to be $0.002 \ {\rm
Mpc}^{-1}$. Chaotic inflation predicts
$r\approx 0.13$ for a $\phi^2$ potential and $r \approx 0.26$ for a
$\phi^4$ potential, and natural inflation predicts $r \approx
0.02-0.05$. 

The potential energy $V$ driving inflation is related to $r$ by $V
\approx (10^{16} \ {\rm Gev})^4 r/0.1$. Low energy inflation models
have low or negligible amplitudes for the CGB. To get the observed
scalar slope and yet small $r$ requires special tuning of the potential.
These are often more complicated, multiple field models, e.g.,
\citep{linde:1993}, or string-inspired brane or moduli
models~\citep{kallosh2007}.  Given the collection of models it is difficult to 
predict a precise range for the expected tensor level and the prior
probability for $r$ should be considered as wide open.

Recent CMB data have reached the sensitivity level required to constrain 
the amplitude of and possibly characterize the gradient-like, or 
E-mode, component of the polarization 
\citep{Kovac:2002fg,hedman2002,Readhead:2004xg,Montroy:2006,page2007,quad2007}.  A significant 
complication of the measurements is that the E-mode amplitude is an 
order of magnitude lower than the total intensity. In addition, galactic 
foregrounds such as synchrotron and dust are expected to be 
significant at these amplitudes~\citep{kogut2007}. Furthermore the 
polarization properties of foregrounds are largely unknown.  Constraining B-modes 
presents an even greater challenge as it is a near certainty that 
polarized foregrounds will dominate the signal. 

The next generation of CMB experiments will benefit from a revolution in
detector fabrication in the form of arrays of antenna-coupled
bolometers~\citep{Goldin:2002,kuo2006}. The antenna-coupled design is
entirely photo-lithographically fabricated, greatly simplifying detector
production.  In addition, the densely populated antennas allow a very
efficient use of the focal plane area.

\spider\ will make use of this technological advance, in the form of 
2624 polarization sensitive detectors observing in four frequency 
bands from $96$ to $275$ GHz.  A multi-frequency observing strategy is a 
necessary requirement to allow for a subtraction of the foreground 
signal. \spider\ will observe over a large 
fraction of the sky at degree scale resolution producing high 
signal-to-noise polarization maps of the foregrounds at each 
frequency. 

Extraordinarily precise understanding of systematic effects within the
telescope will be required to measure the tiny B-mode signal.
This {\it paper} presents a detailed 
investigation of experimental effects which may impact \spider's 
measurement of B-modes.   
The strategy is to simulate a \spider\ flight 
time-stream injecting systematic effects in the time 
domain. The aim is to determine the level of B-mode contamination  
at subsequent stages of the analysis.  With these results one can set 
stringent requirements on experimental design criteria, in addition to 
optimizing the telescope's observing strategy.   
 
The outline of this paper is as follows. 
Section~\ref{sec:instrument} gives an overview of 
the instrument, flight and observing strategy. 
Section~\ref{sec:method} describes the details of the simulation 
methodology.  Results for several systematic effects are presented 
in Section~\ref{sec:results}.  Section~\ref{sec:conclusions} concludes with a 
summary and discussion of the results.

\section{The Instrument}\label{sec:instrument}
\begin{figure*}[th]
\centering
\includegraphics[width=15cm,clip]{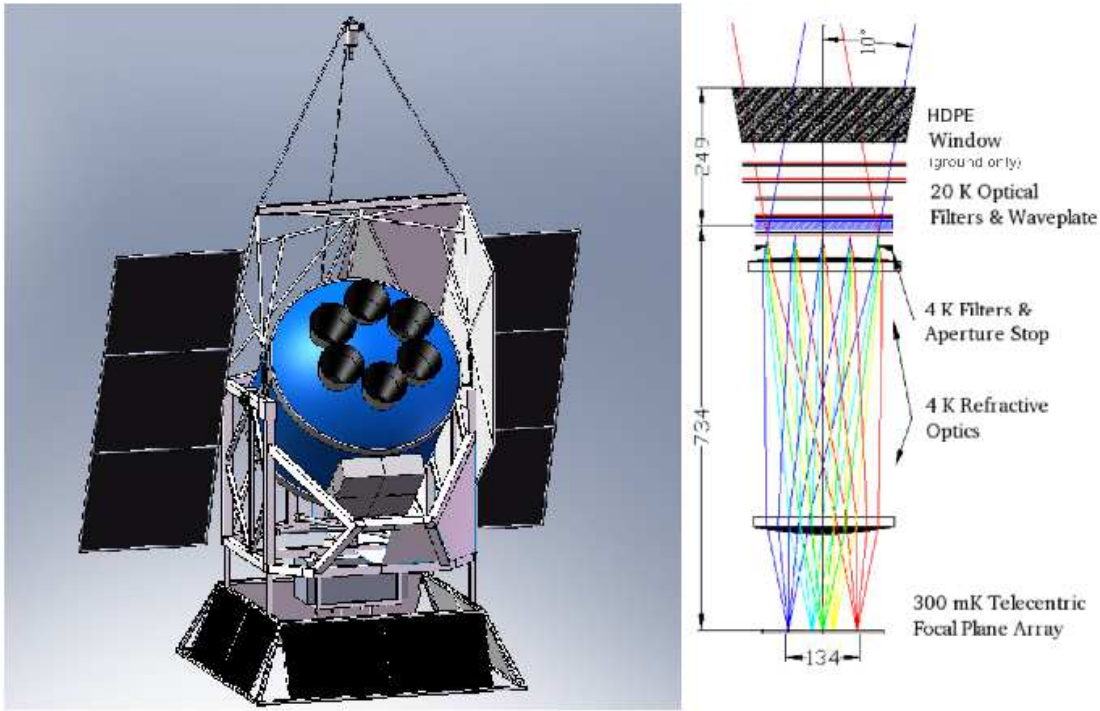}
\caption{\small 
{\em Left:} The \spider\ payload.  Six independent monochromatic
telescopes are housed in a single long hold time cryostat.  Each
telescope is fully baffled from radiation from the ground.  Power is
supplied by solar arrays.  The baseline observing strategy is to spin
the payload in azimuth at fixed elevation.  \spider\ is designed to
obtain maximum sky coverage during a 20-30 day, mid-latitude,
around-the-world flight. {\em Right:} \spider\ optical train.  The
telescope yields a flat and telecentric focal plane.  The apodized
Lyot-stop, which is fixed with regard to the instrument, is maintained
at 4 Kelvin.  All dimensions are in millimeters.}\label{fig:payload}
\end{figure*}

An initial description of \spider\ can be found in~\cite{Montroy:2006a}.
Since that publication some of the telescope features have been changed 
in order to simplify design and to further optimize the instrument.  
An overview of \spider\ instrumentation is given here.  

\spider\ is a balloon-borne polarimeter designed to measure
the polarization of the CMB at large angular scales. Beyond the ability
to map large scales, a clear advantage of a balloon platform is the
increase in raw sensitivity, especially in the higher frequency
channels, achievable above the Earth's atmosphere.  For the
long-duration balloon (LDB) flight \spider\ will launch from Australia,
with a $\sim$25 day, around the world, constant latitude, trajectory.
The first test flight is scheduled for fall 2009, a turn-around flight
from Alice Springs, Australia, of $\sim 48$ hours duration.

A schematic of the \spider\ payload is depicted in the left panel of
Figure~\ref{fig:payload}.  The \spider\ gondola will spin in azimuth at
a fixed elevation, observing only when the sun is a few degrees below
the horizon~\footnote{An additional daytime (anti-sun) scanning mode may
be implemented but is not discussed here.}.  A constant latitude
25-day flight, launching from Australia (with the optical axis tilted at
41 degrees from the Zenith), yields a sky coverage of
$\sim 60$\%.

Azimuthal attitude control is provided by a reaction wheel below the
payload and by a torque motor in the pivot located above the gondola.
\spider\ will employ a
number of sensors to obtain both short and long time scale
pointing solutions. These include:  two star cameras, 3 gyroscopes, a
GPS and a three axis magnetometer.  The pointing system is based on proven
\boomerang\ \citep{Masi:2006} and \blast\ \citep{Pascale:2007} techniques.
The pair of star cameras are mounted above the
cryostat on a rotating platform, which will allow them to remain fixed on
the sky, providing pointing reconstruction accurate to $\sim6''$.
Solar arrays pointing toward the sun during daytime operation will
recharge the batteries supplying payload power.

The instrument consists of 6 monochromatic telescopes operating from
$96$ GHz up to $275$ GHz.  All six telescope inserts are housed in a
single LHe cryostat which provides $>$30 days of cooling power at 4K
(for the optics) and at 1.5K (for the sub-K cooler).  The detectors are
further cooled to 300mK using simple $^3$He closed cycle sorption fridges, one
per insert, which are cycled each day when the sun prevents
observations.  Specifications for each of the 6 telescopes, including
observing bands and detector sensitivities are given in
Table~\ref{tab:specs}.

\begin{figure*}[th]
\centering
\includegraphics[width=15cm,clip]{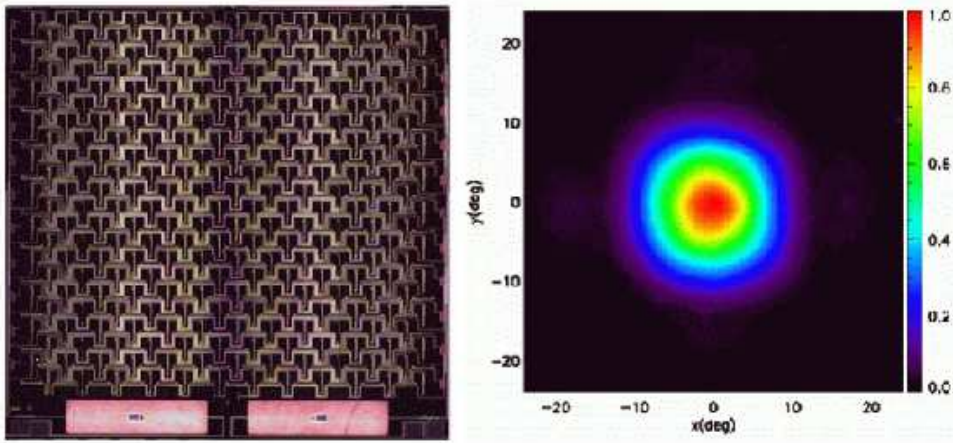}
\caption{\small 
{\em Left:} A single pixel of a 145~GHz antenna-coupled bolometer,
comprising a 288 element phased array
of dual-polarization slot antennas coupled to a matched load by a
superconducting microstrip network.  Microstrip filters, which determine
the spectral response, and TES detectors, which measure the power
dissipated in the load are visible at bottom. {\em Right:} The measured
beam pattern of the dual-polarization antenna.  The upper
limit on differential beam ellipticity of is 1\%, limited by the
testbed. The polarization efficiency is greater than 98\%.
It is important to note that the beam pattern here is the {\it feed} beam
pattern.  The beam on the sky is influenced by the telescope.  While the
\spider\ telescope edge taper is modest, the beam on the sky will be more symmetric than the
feed pattern shown here.  In particular, the visibly large and
asymmetric lobes above will not propagate to the sky.
}\label{fig:acbs}
\end{figure*}

\spider\ uses antenna-coupled bolometer arrays cooled to
300 mK~\citep{kuo2006}.  Figure~\ref{fig:acbs} shows an image of a
prototype detector and the measured beam response of a single
dual-polarization antenna.  The antenna arrays are intrinsically
polarization sensitive, with highly symmetrical beams on the sky and low
sidelobes.  Each spatial pixel consists of phased array of 288 slot
dipole antennas, with a radiation pattern defined by the coherent
interference of the antennas elements.  
Each of the feed antennas provides an edge taper of roughly $-13\pm 1$
dB on the primary aperture.  A single spatial pixel has orthogonally
polarized antennas.  The optical power incident on an antenna is
transmitted to a bolometer and detected with a superconducting
transition-edge sensor (TES) immediately adjacent to the spatial pixel.

The TES sensors will be read out using superconducting quantum
interference device (SQUID) current amplifiers with time-domain
multiplexing~\citep{1999chervenak, 2003dekorte, 2003reintsema,
2004irwin}.  Ambient temperature multi-channel electronics (MCE) \citep{Battistelli:2007},
initially developed for SCUBA2~\citep{2006holland}, will work in concert with
the time-domain multiplexers.

The optical design for the inserts, shown in the right panel of
Figure~\ref{fig:payload}, is based on the \bicep \citep{keating2003}
optics.  The monochromatic, telecentric refractor comprises two
AR-coated polyethylene lenses and is cooled to 4K in order to reduce the
instrumental background to negligible levels.  The
primary optic is 302 mm in diameter and the clear aperture of the
Lyot-stop is 264 mm which produces a 45' beam at 145 GHz.  

\begin{table*}
\begin{center}
\begin{tabular}{|l|c|c|c|c|c|c|}
\hline
Obs. Band (GHz) & 96 &96 &145 &145 &225 &275\\ 
\hline\hline
Orientation & Q & U & Q & U & Q & U \\
Bandwidth (GHz) & 24 & 24 & 35 & 35 & 54 & 66 \\ 
Number of Detectors & 288 &288&512&512&512&512\\
NET ($\mu$K$\sqrt{\mbox{s}}$)&100&100&100&100&204&351\\
Beam FWHM (arcmin) &58&58&40&40&26&21\\
\hline
\end{tabular}
\end{center}
\caption{{\rm \spider\ Channel Specifications.  Instrument orientation, observing bands, detector
counts, sensitivities and beams.  A total of 2624 detectors is
distributed between six telescopes, with two operating at 96 GHz and
two operating at 145 GHz.}}\label{tab:specs} 
\end{table*}

A cryogenic half-wave plate is located in front of the Lyot-stop of
each telescope.  Rotating the half-wave plate aids in polarization modulation,
making for cleaner measurements of the Stokes Q and U, mitigating the 
need to difference detector time-streams.  
This is essential to the reduction of the requirements for
characterization of individual detectors and ultimately a reduction of experimental
systematics.  The half-wave plate consists of a
single birefringent sapphire plate coated with a single layer of Herasil
quartz on each side.  

Initially polarization modulation was to be achieved via a {\it continuously spinning
half-wave plate}~\citep{Montroy:2006a}.  This work examines the viability
of a {\it fast spinning gondola} modulating the incoming signal with the
half-wave plate stepping 22.5 degrees per day.  Section~\ref{sec:modes}
illustrates that either of these modes of operation can be used for \spider.
The latter mode, stepping the half-wave plate, is easier to design mechanically 
and more robust to operate, and is therefore preferred. 

\section{Simulation Methodology}\label{sec:method}

\begin{figure*}
\centering
\includegraphics[width=15cm,clip]{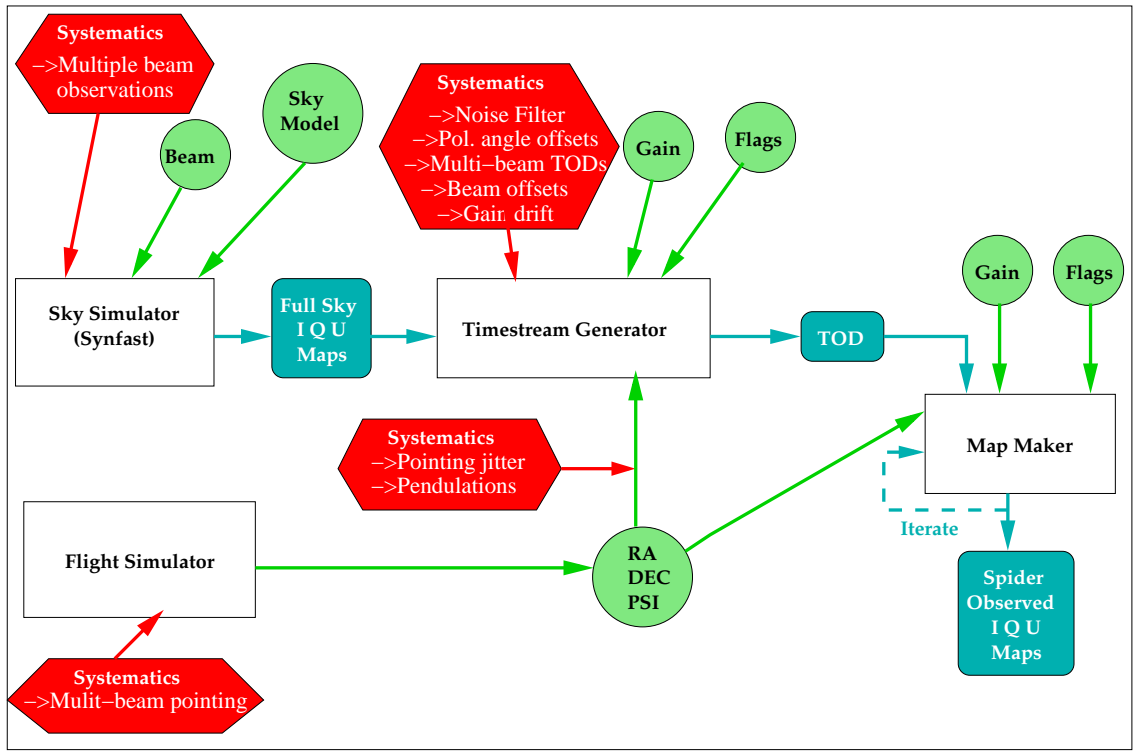}
\caption{\small A schematic representation of the simulation pipeline.
  During the time-stream generation
 the (stepped or spinning) half-wave plate polarization angle is
 added to the intrinsic polarization angle of the individual detectors.
 In addition polarization angle systematics, beam offsets and gain drift are also 
 applied during time-stream generation.  Pointing jitter and
 pendulation systematics are added to the pointing time-streams during
 flight simulation.  For the case of multiple beam distortions, multiple
 pointing time-streams are produced and the full-sky map
 is observed by each beam/pointing.  
}\label{fig:pipeline}
\end{figure*}

The simulation pipeline is based largely on the analysis pipeline
described in~\cite{Jones:2007} which was
developed for the analysis of the data obtained from observations made with
the \boomerang\ telescope during the 2003 LDB Antarctic
flight~\citep{Montroy:2006,Piacentini:2005yq,Jones:2005yb,Masi:2006}.
A schematic outlining the components of the simulation pipeline and the
various inputs and outputs is given in Figure~\ref{fig:pipeline}.

The flight simulator generates time-ordered pointing data in the form
of right ascension, declination and polarization angle for each
detector.  Data are simulated for 16 detectors, arranged in evenly spaced
pairs in a single column which extends the full height of the focal
plane.  Detectors in a pair are oriented to be sensitive to orthogonal
polarizations.  Since signal-only simulations are used in this work
this is sufficient to test most of the systematic effects and
observing strategies considered here. The small number of time-streams
also reduces significantly the data storage and computation
requirements which itself will present a unique challenge for
the actual analysis.

It is assumed that the telescope is fixed in elevation
41 degrees from the zenith and that the payload is moving in longitude (beginning
at 128.5 east) at a speed of $3.76\times 10^{-4}$ degrees per second at 
constant latitude (25.5 south).  Data are simulated for four days of
operation, assuming a mid-November launch.  Four days operation allows
for one complete observing cycle for the stepped half-wave plate operating
mode, after which the cycle is repeated.  This is also the minimum required to
ensure sufficient coverage for polarization reconstruction of the
entire observed area. 

Full sky intensity and polarization maps are simulated and smoothed with
the \spider\ beam using the {\it synfast} program which is part of the
{\rm HEALPix} software~\citep{Gorski:2005}.  In order to ensure that signal
variation within a pixel is negligibly small the full sky maps are
pixelized at a resolution which corresponds to a pixel size of
$\sim3.4'$.  

Full sky maps are then converted into time-ordered data
(TOD) using pointing information from the flight simulator.  Thus, the
time-stream generator constructs $d_i$ for each detector from the
equation
\begin{equation}\label{eq:tod}
        d_i = G[I_{pix} + \frac{\rho}{2-\rho}(Q_{pix} {\rm cos}(2\psi_i)
+ U_{pix} {\rm sin}(2\psi_i))]. 
\end{equation}
Here $\rho$ parametrizes the polarization
efficiency, $\psi$ is the final projection of the orientation of a
detector on the sky and $G$ is the detector gain or responsivity.  

All time-streams are high-pass filtered at $10$ mHz during the
map-making stage.  This is done to test the impact of the filtering that
is required in the case of real data which is effected by long-timescale
systematics.  Particular systematics of concern are knowledge of system
transfer functions (or equivalently knowledge of the gains) and
knowledge of the noise amplitude/statistics on long-timescales.

During the time-stream generation
the (stepped or spinning) half-wave plate polarization angle is
added to the intrinsic polarization angle of the individual detectors.
In addition polarization angle systematics, beam offsets and gain drift are also 
applied during time-stream generation.  Additional pointing jitter and
pendulation systematics are added to the pointing time-streams during
flight simulation.  For the case of multiple beam distortions, multiple
pointing time-streams are produced and the full-sky map
is observed by each beam.  

Finally, \spider\ maps are constructed in terms of the observed
Stokes parameters, $I^{\rm obs}$, $Q^{\rm obs}$, and $U^{\rm obs}$ with
an iterative map maker that uses an adaptation of the Jacobi method
(described in detail in~\cite{Jones:2007}) to recover the input signal.
To reduce computation time output maps are pixelized at a resolution
which corresponds to a pixel size of $\sim13.7'$ (about 1/4 of the beam
size). Although the simulations are for pure signal, the map maker
algorithm performs an inverse noise filtering of the time-streams.  This
filtering would be included in noisy time-streams to reduce the
strongest effects of $1/f$ noise which can significantly reduce
map-making efficiency.  In this case it is
also included to make the simulations used here accurate representations
of the full pipeline.

\begin{figure*}[th]
\centering
\includegraphics[width=17cm,clip]{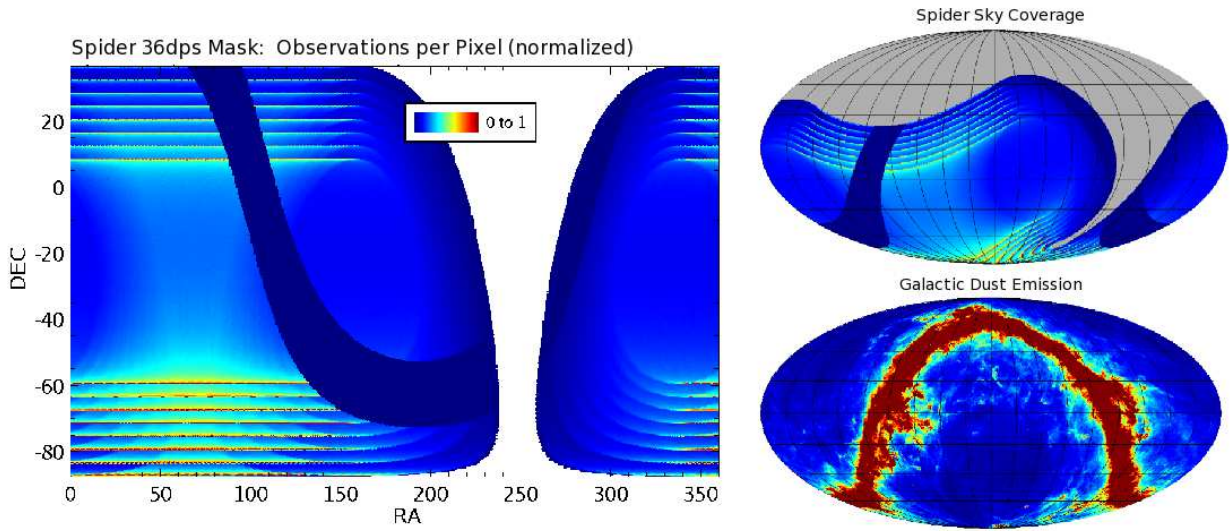}
\caption{\small 
{\em Left:} The \spider\ mask for 36 dps simulations.  The dark band
represents a galactic cut at $\pm 10^{\circ}$ in galactic latitude and
the white region represents sun flagging.  With these regions flagged
the fraction of the sky covered for this observing strategy is $\sim 60$\%.
Pixel values are the number of observations in the pixel divided by the
maximum hits value.  The most obvious features are constant declination
lines where scan circles on the sky for each detector overlap and the
coverage is deepest. The pixel weighting is applied in order to reduce
the effect of badly sampled pixels at the edge of the map.  {\em Top
Right:} \spider\ coverage projected into equatorial coordinates.  {\em
Bottom Right:} The IRAS 100 $\mu$m map~\citep{SFD:1998} of Galactic dust
is shown for comparison.  }\label{fig:mask}
\end{figure*}

\subsection{Residual measure}

The aim is to quantify the contamination of the B-mode angular power
spectrum (BB) from systematics which induce either ${\rm I}  \rightarrow
{\rm Q, U}$ or ${\rm Q} \leftrightarrow {\rm U}$ mixing.  To assess the impact
of the various systematics on BB the following procedure is
implemented:

\begin{itemize}
\item Generate full-sky input $I$, $Q$, and $U$ maps with $C_\ell^{BB}=0$.
\item 'Observe' the maps with simulation pipeline including a chosen
  systematic (but no noise) giving signal only $I^{\rm obs}$, $Q^{\rm
  obs}$, and $U^{\rm obs}$.
\item Take the difference between input and output maps over the survey area
\begin{eqnarray}
  I^{\rm res} &=&  I^{\rm obs}-I,\nonumber\\
  Q^{\rm res} &=&  Q^{\rm obs}-Q,\\
  U^{\rm res} &=&  U^{\rm obs}-U.\nonumber
\end{eqnarray}
\item Apply a mask with pixel weighting determined by the number of observations per pixel
to the residual maps.
\item Spherical harmonic transform the weighted maps to obtain
pseudo-$C_\ell$ spectra of the BB residual.
\item Compute residual measure $R^{BB}_\ell$ defined below.
\end{itemize}

The pixel weighting is applied in order to reduce the effect of badly
sampled pixels at the edge of the map.  These would bias heavily the
raw pseudo-$C_\ell$ computed from the residual maps.  The mask that is applied to all
of the simulations for which the gondola spin rate is 36 degrees per
second (dps) is shown in Figure~\ref{fig:mask}.  The dark band
represents a galactic cut at $\pm 10^{\circ}$ in galactic
latitude.  The white region represents the portion of the sky that
cannot be observed because of the sun.  With these regions flagged the
fraction of the sky covered for this observing strategy is
$\sim 60$\%.  Each pixel value in the mask is the number
of observations in the pixel divided by the value in the pixel with the 
maximum hits.

The spectra obtained from the method above are raw cut-sky, or
pseudo-$C_\ell$, power spectra \citep{Hivon:2002}. Since no B-mode power is
present in the original full-sky simulation any B-mode power in the
final maps will be due to the mixing of modes from either
systematics or cut-sky effects which mix E and B-modes
\citep{Lewis:2002,Bunn:2003}. In the limit of full-sky a simple model for the
spherical harmonic coefficients of the residual maps is given by
\begin{equation}\label{eq:alm}
  a^{Bres\,(\rm noBB)}_{\ell m} =\left[F_\ell^{E\rightarrow
  B}\right]^{1/2} a_{\ell m}^{E}+ \left[F_\ell^{T\rightarrow B}\right]^{1/2} a_{\ell m}^{T}.
\end{equation}
The terms on the right hand side of Eq.~\ref{eq:alm} represent 
leakage of total intensity and E-mode into B due to time domain
effects induced by the pipeline. They include any loss of modes
due to time domain filtering. The total transfer is described as an
isotropic coefficient $F_\ell$. There is no $E\leftrightarrow B$
mixing due to any cuts in this case.

A significant assumption introduced above is that time domain
effects result in isotropized transfer of mode power in the map
domain, hence no $m$-dependence of the transfer functions
\citep{Hivon:2002}. The validity of the assumption depends on the observing
strategy adopted but should be appropriate for polarization
experiments where cross-linking is maximized.

When including cut-sky effects the pseudo-$C_\ell$ residual B-mode spectrum
can then be approximated as
\begin{eqnarray}\label{eq:res}
  \widetilde C_\ell^{BBres\,(\rm noBB)} &=& \ \ \sum_{\ell'}
{}_{-}K^{\,}_{\ell\ell'}B_{\ell'}^2\left(\left[F_{\ell'}^{(1)}\right]^{1/2}-1\right)^2C_\ell^{EE}\nonumber\\
  &&+\sum_{\ell'} M^{\,}_{\ell\ell'}B^2_{\ell'}F_{\ell'}^{(2)}
  C_{\ell'}^{TT} \nonumber\\&&-2\sum_{\ell'}{}_{\times}K^{\,}_{\ell\ell'}B^2_{\ell'}F_{\ell'}^{(3)}C_{\ell'}^{TE}.
\end{eqnarray}
Here the $M_{\ell\ell'}$ is the geometric total intensity kernel,
${}_{-}K^{\,}_{\ell\ell'}$ is the geometric leakage kernel
\citep{Spice:2001} coupling $E\rightarrow B$, and
${}_{\times}K^{\,}_{\ell\ell'}$ is the geometric kernel for the
cross-correlation spectrum. The transfer functions
$F_{\ell'}^{(1,2,3)}$ represent a combination of individual transfer
  effects
\begin{eqnarray}
  F_\ell^{(1)} &=& F_\ell^E+F_\ell^{E\rightarrow
  B}+2\left(F_\ell^EF_\ell^{E\rightarrow B}\right)^{1/2},\\
  F_\ell^{(2)} &=& F_\ell^{T\rightarrow E},\\
  F_\ell^{(3)} &=& \left(\left[F_\ell^E\right] + \left[F_\ell^{E\rightarrow
B}\right]^{1/2}\right)^{1/2}\left[F_\ell^{T\rightarrow B}\right]^{1/2}.
\end{eqnarray}

The quantity, $\widetilde C_\ell^{BBres\,(\rm noBB)}$ in (\ref{eq:res}) is divided by $\widetilde{C}^{BB (no EE)}_{\ell}$
i.e. the BB signal obtained from reconstructed \spider\ Q/U maps with
only an input BB signal (for r = 0.1) and no input EE
\begin{equation}\label{eq:noee}
\widetilde{C}^{BB (no EE)}_{\ell} =
\sum_{\ell'}{_{+}K_{\ell\ell'}B^2_{\ell'}F^{BB}_{\ell'}C^{BB}_{\ell'}}.
\end{equation}
Dividing (\ref{eq:res}) by (\ref{eq:noee}) gives the fraction of the
raw BB power that is coming from a transfer effect caused by systematics
and not from an input BB spectrum.
The same mask, with pixel weighting determined by the number of
observations per pixel, is applied to {\it all} maps (residual and Q/U)
which have the same gondola spin rate.  The final BB
residual measure is then defined as
\begin{equation}\label{eq:measure}
   {R}^{BB}_{\ell} = \frac{ \widetilde{C}^{BBres\,(\rm
  no BB)}_{\ell}}{\widetilde{C}^{BB\,({\rm no
  EE})}_{\ell}}C^{BB}_{\ell}.
\end{equation}
This multiplies the fractional residual by the input BB spectrum for r =
0.1 giving the residual in terms of an equivalent BB signal.
The residual measure defined above is not designed to give a complete
picture of how well the original BB signal can be reconstructed from
the observations. A complete treatment would require a full un-biased
power spectrum estimation method, which is beyond the scope of this
work. Instead~(\ref{eq:measure}) isolates the impact of the systematics
under study on the observed signal by minimizing the impact of the
$E\rightarrow B$ mixing from cut-sky effects.

Note that for all of the input maps the same initial seed value is used
to generate the full CMB sky, i.e. the sample scatter is the same for
all simulations.  

\section{Simulation Results}\label{sec:results}

The presentation of results begins by illustrating the base residuals for two basic modes of
half-wave plate operation--stepped and continuously rotating.  For the
remaining subsections the B-mode residuals from experimental systematic
effects for the stepped half-wave plate case are examined.  All
simulations are for signal only (with no noise) but time-streams are
inverse noise filtered during the map making phase.  Aside from
Section~\ref{sec:noise}, which explores two knee frequency values, the
1/f knee for the noise filter is 100 mHz for all simulations.  In all
plots the case labeled {\it nominal} is a 36 dps gondola spin rate, with
the half-wave plate stepping $22.5^{\circ}$ once per day, with 10
iterations (sufficient to recover the residual levels of the
continuously-rotating half-wave plate case) of the map-maker, a Jacobi iterative
solver~\citep{Jones:2007}.

\subsection{Polarization Modulation}\label{sec:modes}

\begin{figure*}[th]
\centering
\includegraphics[width=12cm]{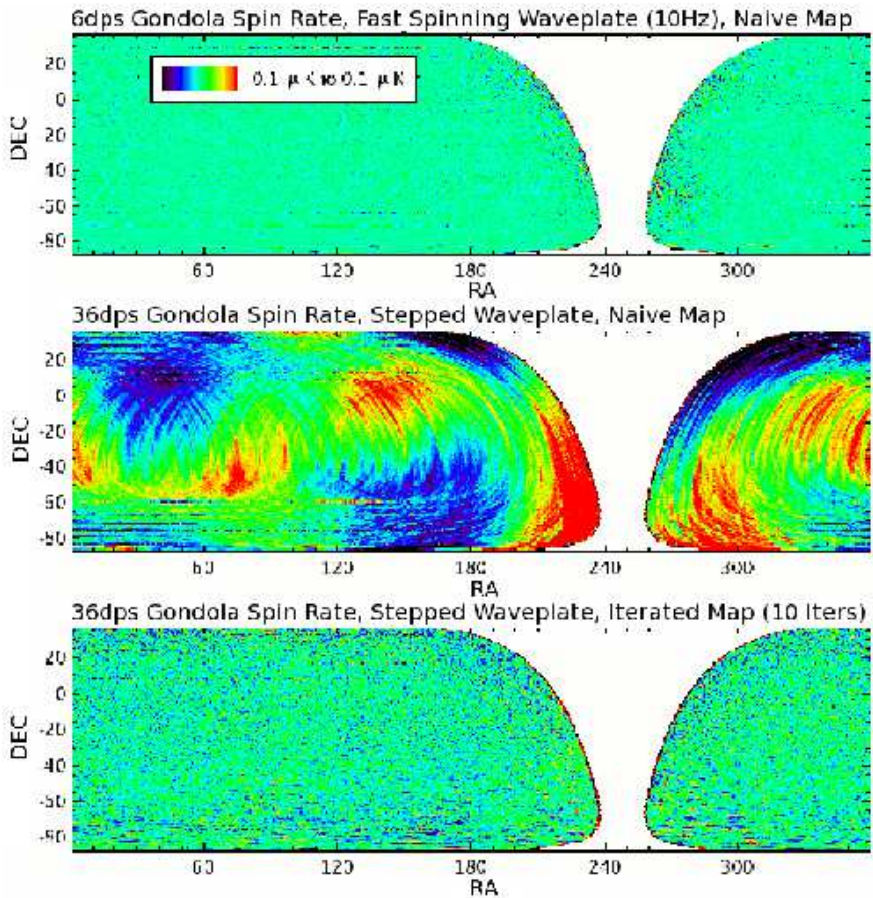}
\caption{\small Maps of the residuals in the Q Stokes
parameters for the two polarization modulation strategies. The
top panel shows the residuals for the continuous
half-wave plate rotation case. For this case a ``naive'', or
zero-iteration, map is shown. The middle panel shows the naive map
for the stepped half-wave plate mode with the gondola spinning at 36 dps. In
this case significant striping is present due to the loss of low
frequency modes.  This is caused by the inverse noise filtering of the 
time-streams during the map-making phase which uses a noise kernel with a
realistic 100 mHz 1/f knee.  For the stepped case the polarization
modulation is not sufficient.  However, iterated map-making reduces the impact of the
striping as shown in the bottom panel and 10 iterations of the
map-maker are sufficient to recover most of the lost modes. 
}\label{fig:resmaps}
\end{figure*}

\begin{figure*}
\centering
\includegraphics[width=10cm,clip,angle=-90]{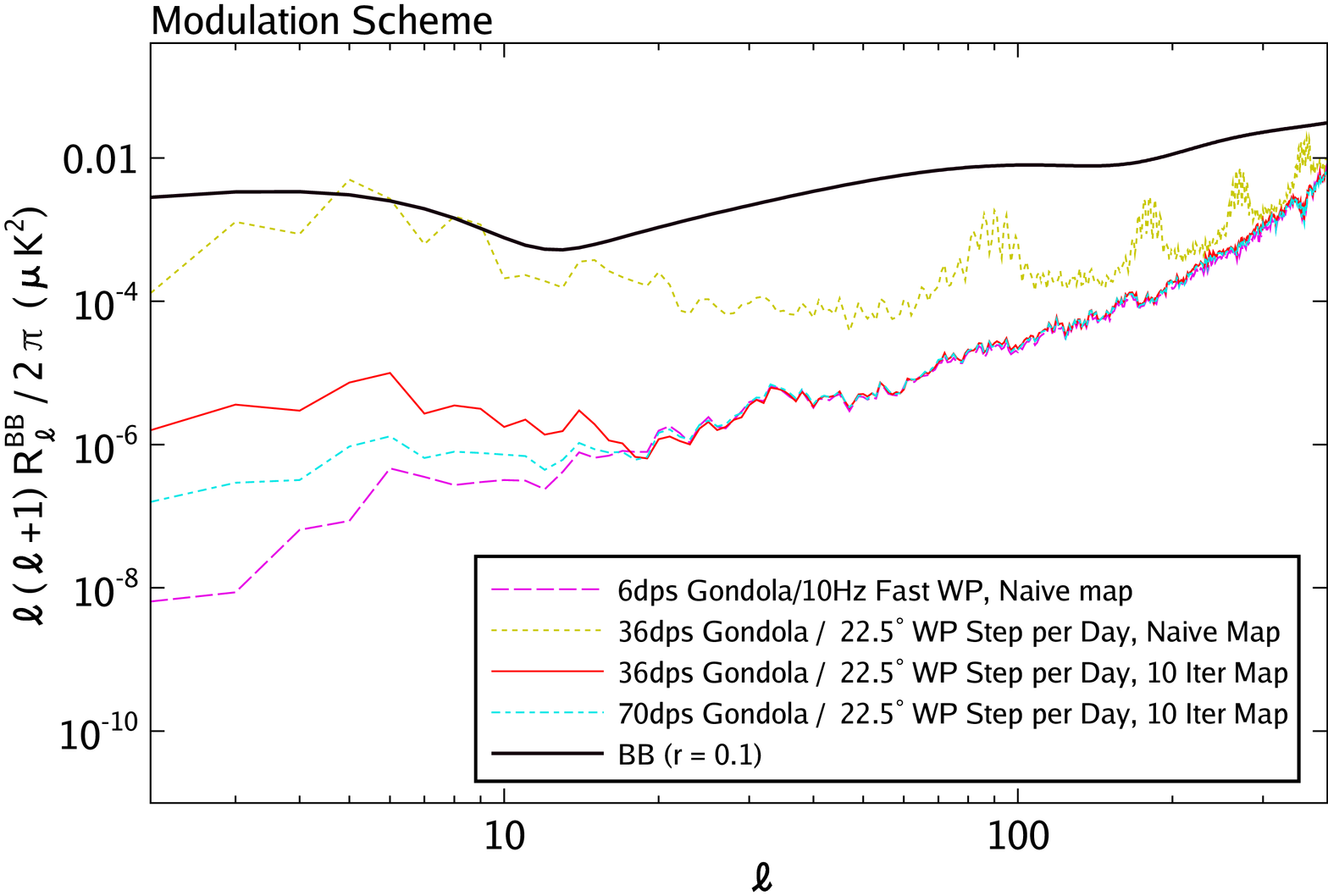}
\caption{\small Comparison of the BB residual from
   continuously spinning and stepped half-wave plate
  polarization modulation schemes. For the
first case the half-wave plate spins continuously at 10 Hz while the gondola
rotates at 6 dps. For the second case the half-wave plate is
stepped by 22.5 degrees per day, while the gondola rotates at 36 dps or more. 
  Operation in stepped half-wave plate mode,
  with the modulation provided by the spinning gondola, requires iterated
  map-making as the signal is not modulated as far from the $1/f$ knee
  as in the continuously-rotating half-wave plate mode. 10 iterations of
  the map-maker is sufficient to
  recover the residual levels of the continuously-rotating half-wave plate case. Fewer
  iterations may be required if the gondola spin rate is even higher (70
dps case).  The input BB spectrum for r = 0.1 is plotted for comparison.
}\label{fig:fast}
\end{figure*}

Since \spider's default observing strategy is to spin continuously in
azimuth, two modes of polarization modulation are explored. For the
first case the half-wave plate spins continuously at 10 Hz while the gondola
rotates at 6 dps. For the second case the half-wave plate is
stepped by 22.5 degrees per day, while the gondola rotates at 36dps or more.  Therefore in the first instance the
half-wave plate is modulating the incoming polarization signal and in the
second case the gondola itself is used to modulate the
signal.

Modulation by the gondola spin has a number of design advantages over
the inclusion, in the optical train, of a continuously rotating
half-wave plate. The half-wave plate adds a degree of complexity in the design
with a subsequent impact on the robustness of the instrument. In
addition it is a potential source of a number of systematic
effects for example microphonics, thermalization effects, magnetic pickup
and higher power dissipation at 4K.  It is
therefore preferable (and nearly equivalent as will be shown) to
step the half-wave plate once per day, in order to increase Q and U redundancy in a single pixel,
while rapidly spinning the gondola in order to move the signal above the detector
$1/f$ knee frequency.  

Figure~\ref{fig:resmaps} shows Q residual maps for the two modulation
modes.  Maps of the U residuals are not shown here but are of similar
amplitude. The top panel shows the residuals for the first case
(continuous half-wave plate rotation at 10Hz). For this case a ``naive'', or
zero-iteration, map is shown. The naive map is equivalent to a simple
(pixel-hit weighted) binning of the time-stream into pixels. Iterations of
the map-making step are not required in this case since the signal is
modulated to frequencies higher than the expected $1/f$ knee and the
loss of modes at low frequencies has virtually no impact on the final
maps. This is one of the benefits of a design which includes a
continuously rotating half-wave plate.

The middle panel of Figure~\ref{fig:resmaps} shows the naive map
for the stepped half-wave plate mode with the gondola spinning at 36 dps. In
this case significant striping is present due to the loss of low
frequency modes.  These translate to large scale modes along the
individual scans and result in the striping obvious in the maps.
Iterating the map-maker in this case reduces the effect of striping as
the large scale modes are recovered. After 10 iterations the striping
is significantly reduced as shown in the bottom panel of
Figure~\ref{fig:resmaps}. One possible way to reduce the computational
load of a map-making stage with many iterations is to spin the gondola
faster to modulate the signal into higher frequencies. 

The power spectra for the BB residual measure (\ref{eq:measure}) are
shown in Figure~\ref{fig:fast}. The optimal solution is the continuous
half-wave plate modulation scheme. This yields the lowest residual compared
to an $r=0.1$ fiducial BB model. In the stepped the 36 dps,
non-iterated case the residuals have the same amplitude as the model
on the largest scales. The residuals are reduced to $< 1$\% levels
for multipoles $\ell < 100$ when the map-maker is iterated 10
times. The stepped 70 dps spin case with 10 iterations yields
even smaller residuals at the largest scales.  

Given the design and implementation advantages, the simple stepped
half-wave plate system appears a feasible choice for the \spider\ scan
strategy, albeit with significant additional computational costs
\footnote{A full exploration of faster, sub-optimal map-making
  algorithms is left for future work. In particular de-striping
  algorithms (see e.g.~\cite{Ashdown:2007}) may provide a much faster alternative although it is
  still not clear that these can be applied to a \spider\
  observing strategy and polarization sensitivity requirements.}. The
  remainder of this Section is restricted to the stepped half-wave plate
  case. In particular the focus will be to probe whether any other
  systematic effects invalidate this choice of modulation scheme.

\subsection{Noise}\label{sec:noise}

To examine the impact of different $1/f$ profiles on the stepped mode
residuals a number of different cases were run.
\begin{itemize}
\item 36 dps gondola spin rate with 100 mHz
detector $1/f$ knee.
\item 36 dps gondola spin rate with 500 mHz
$1/f$ knee.
\item 110 dps gondola spin rate with 500 mHz $1/f$
knee.
\end{itemize}

\begin{figure*}
\centering \includegraphics[width=10cm,clip,angle=-90]{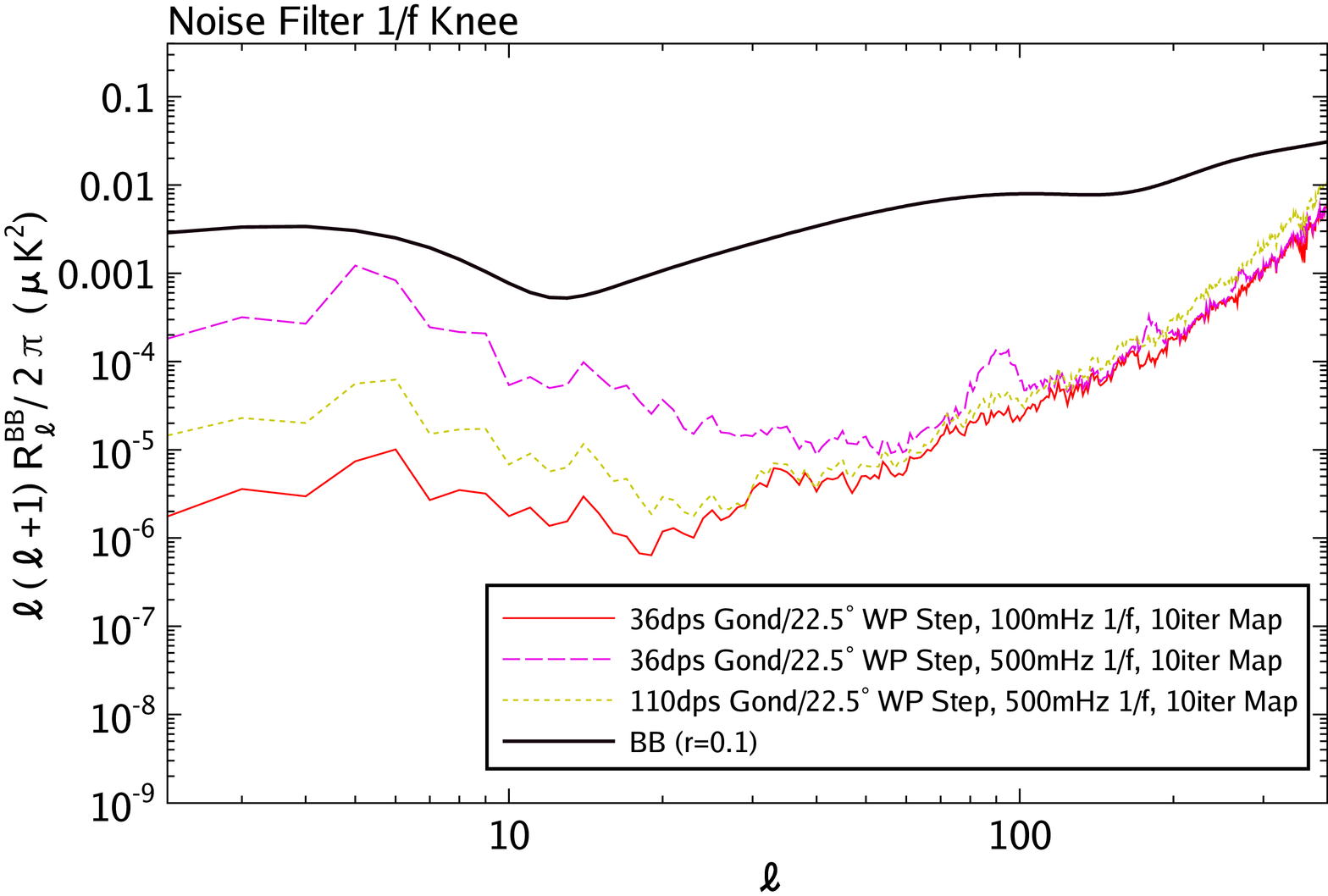}
\caption{\small Impact of a higher $1/f$ detector knee on the BB
  residuals. Simulations are for signal only (with no noise) but 
  time-streams are inverse noise filtered in the map-maker, giving a
  realistic estimate of modes that would be lost from 1/f effects 
  in the real data.  With the higher knee frequency of 500 mHz 
  the gondola spin rate must be increased to reduce the
  residuals to $<$2\% levels over the target range of multipoles
  $\ell < 100$.  The maximum gondola spin rate of $\sim 110$ dps 
  is limited by the time response and noise characteristics of the
\spider\ detectors. The input BB spectrum for r = 0.1 is plotted for comparison. 
}\label{fig:fknee}
\end{figure*}

The time-stream high-pass filter cut-off is kept at 10 mHz in all cases.
A comparison of BB signal residuals varying the detector knee frequency
is shown in Figure~\ref{fig:fknee}.  Although simulations are signal
only the time-streams are inverse noise filtered, as would be done for
the real data.  This reveals the impact of the detector noise
characteristics in terms of the degradation of the polarization signal
on the largest scales.  The effect of a 500 mHz knee is clearly seen on
the largest angular scales. Even for 10 iterations of the map-maker the
residuals are close to the 20\% level for this case. Increasing the spin
rate to 110 dps reduces the impact of the higher $1/f$ knee and
approaches the nominal 36 dps, 100 mHz knee case.  A 500 mHz knee
frequency for the detectors and readout electronics is pessimistic, but
would not be catastrophic since polarization modulation can still be
achieved by the faster spinning gondola.  \spider's high frequency
response is limited by the noise and response time of the detectors the
combination of which sets the maximum gondola spin rate.
With 5 ms detectors, \spider\ can spin up to 110 dps before being
affected by the detector noise and time constants.
Thus for the stepped half-wave plate case, the limit for the 1/f 
knee frequency is $\sim$500 mHz (for r = 0.1).    

\subsection{Pointing Systematics}\label{sec:point}

One of the more challenging aspects of balloon-borne telescope
observations is pointing reconstruction, constituted as one of the
limiting systematics in the interpretation of CMB data obtained from a
balloon-borne telescopes.  For \spider\ the pointing system is based on
largely on
\boomerang\ \citep{Masi:2006} and \blast\ \citep{Pascale:2007}
instrument pointing.  The
former achieved $\sim 1.5$ arcminute pointing and the latter
reconstructed pointing to a few arcseconds. The requirement for
\spider\ will be much less demanding given the low resolution.
Nonetheless the design goal for \spider\ is sub-arcminute pointing
reconstruction and establishing a precise requirement is still
important given the particular sensitivity of polarization
measurements to offsets in the pointing. 

Two main pointing offsets are explored in this work.  The first is a
random pointing jitter added to the original pointing solution. The
offsets are added to the right ascension (RA) and declination (DEC)
value of each sample in the solution.  The offsets are constant for 6
seconds, after which new random values are drawn.  The 6 second
bandwidth regime is chosen since this will mimic pointing systematics
which occur within the time-scale of one gondola
spin~\footnote{White-noise pointing jitter, of similar amplitude, that
varied from 
sample-to-sample was also explored and was found to have an insignificant effect.}. 
A nominal run with 1 arcminute RMS and a worst case
scenario with 10 arcminute RMS for the instrument jitter are
considered. To translate from instrument jitter to true jitter on the
sky a factor of $cos(\rm DEC)$ is applied to the RA offsets.

The second systematic consists of a
sinusoidal oscillation with an amplitude of 6 arcminutes and a
20 minute period. This effect simulates the pendulation of the
gondola. The pointing offsets examined are typical of in-flight
conditions--albeit the 10 arcminute jitter is extremely pessimistic.

\begin{figure*}
\centering \includegraphics[width=10cm,clip,angle=-90]{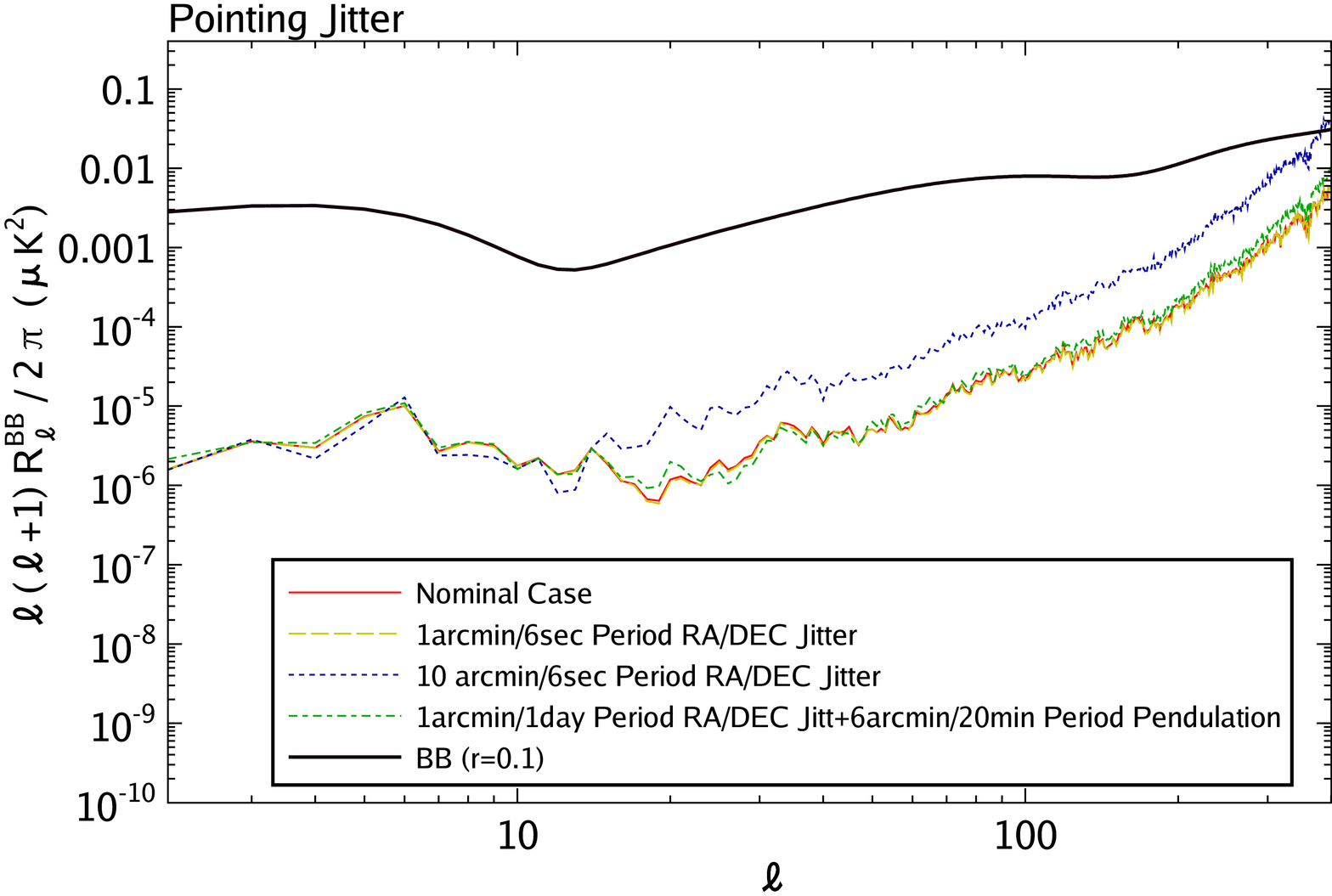}
\caption{\small 
Three cases of pointing jitter.  
In the first two offsets are added to the RA and DEC with new offset
values (with RMS amplitudes 1 arcminute or 10 arcminutes) every 6 seconds. 
In the third case, a sinusoidal 
oscillation is implemented with an amplitude of 6 arcminutes and a 20 minute
period.  This effect simulates the pendulation of the
gondola.  In addition for the latter case, a long time scale (day period) 1 arcminute RMS
jitter is added.  Both types of pointing error, reconstruction error and
in-flight pendulations, have negligible effects; a less than 1\% effect even for
the large 10 arcminute jitter.
}\label{fig:point}
\end{figure*}

The results for the residual measure for the three cases are shown in
Figure~\ref{fig:point}. The pendulation case also includes a
long-timescale 1 arcminute RMS jitter. All cases except the large 10
arcminute RMS jitter result in negligible contributions to the
residual measure at $\ell <100$ compared to the nominal stepped
half-wave plate/36 dps spin mode without any systematic.  

\begin{figure*}
\centering \includegraphics[width=10cm,clip,angle=-90]{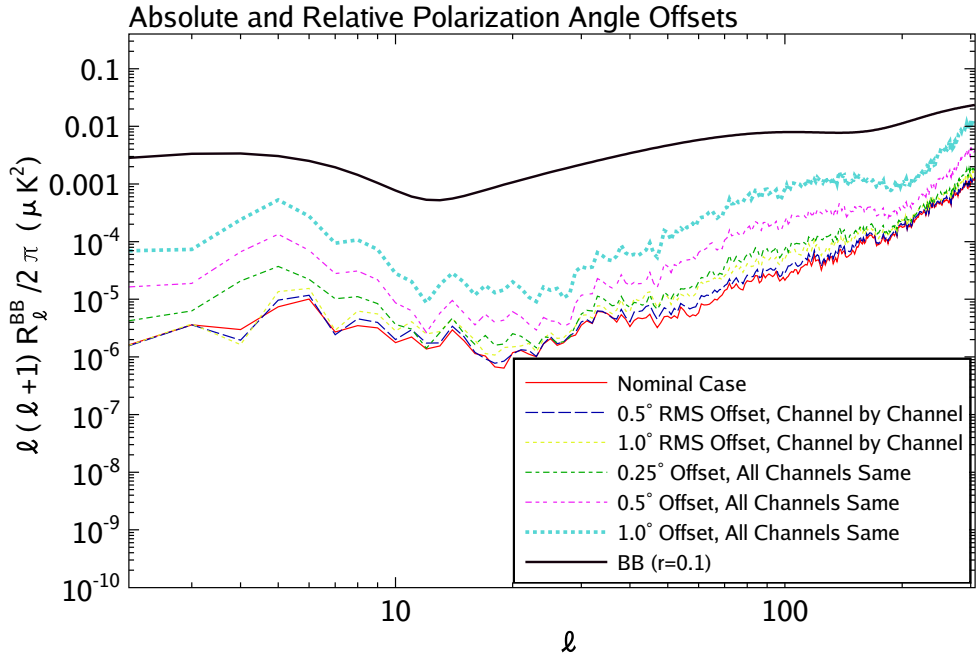}
\caption{\small Residuals from absolute and relative offset of detector
polarization angles.  In the first two cases random 0.5 and 1 degree
RMS $\psi$ errors are added to each detector.  Residuals for these
relative offsets are negligible.  In the remaining cases the same offset 
(0.25 degree, 0.5 degree and 1 degree) is applied to all channels, simulating an overall
calibration error in the half-wave plate $\psi$ angle.  For a 1 degree
absolute offset the effect is as high as 20\%.  The 0.5 degree offsets are required
having only a few percent residual effect.
}\label{fig:psi}
\end{figure*}

An additional pointing systematic affecting polarization measurements
is the requirement to reconstruct the angle $\psi$ (in
equation~\ref{eq:tod}) of the detector polarization
relative to the fixed, local $Q$ and $U$ frame of reference on the
sky.  The systematic can arise in two distinct ways. The first is a
relative offset between the $\psi$ angles of different detectors. The
second is an overall offset in the focal plane reference frame and the
frame on the sky. The latter is generated by any error in the
calibration of the polarization angle of the instrument. 

Results from simulation of the $\psi$ systematics are shown in
Figure~\ref{fig:psi}. In the first two cases random 0.5 and 1 degree
RMS $\psi$ errors are added to each detector. This simulated fixed,
random offsets in the relative polarization angles of the
detectors. The results show the relative offsets contribute a
comparable amount to the residuals as the nominal stepped mode case. 

In the remaining cases the same offset (0.25 degree, 0.5 degree and 1
degree) is applied to all channels. This simulates an overall
calibration error in the half-wave plate $\psi$ angle. The results show
this systematic gives a much larger contribution to the residual.  For a
1 degree absolute offset the effect is as high as 20\%.  The 0.5 degree
offsets is required having only a few percent residual effect.  Again,
simulations consider only 8 pairs in a single column or 16 detectors
total.  The RMS result for the full focal plane should average down as
$\sqrt{N}$, where $N$ is the number of detectors.  This factor has not
been applied to the result.  The calibration of the
half-wave plate $\psi$ angles will be preformed pre-flight on the
ground.  The level of sub-percent precision required is not a difficult
measurement and is made much easier by the compact optics and
correspondingly close ``far field'' of \spider.

\begin{figure*}
\centering \includegraphics[width=10cm,clip,angle=-90]{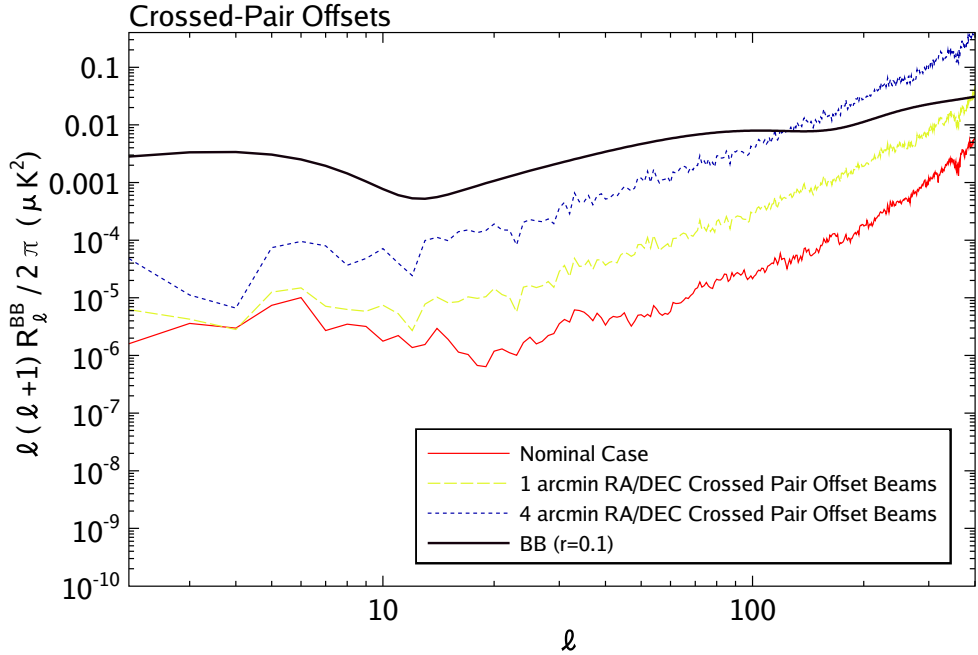}
\caption{\small For these simulations one detector in a
crossed-pair (of the eight pairs) has a constant
offset in RA and DEC of 1 or 4 arcminutes.  For \spider\ 40' beams, the corresponding A-B
amplitudes are 3.6\% and 14\% for 1 arcminute and 4 arcminute offsets. Residuals for beam offsets are plotted in
Figure~\ref{fig:pairoff}.  The effect is less than a few percent for $\ell < 80$ for the largest offset case.
}\label{fig:pairoff}
\end{figure*}

Another systematic that is tested concerns pointing offsets of crossed
pairs.  This occurs if the E-field distribution is not identical at the
feed, or if there are polarization dependent properties in the optics.
For these simulations one detector in a crossed pair (of the eight
pairs) has a constant offset in RA and DEC of 1 or 4 arcminutes.  Again,
for the case of RA a factor of $cos(\rm DEC)$ is applied to the offset and
hence represents the true offset on the sky.  For \spider\ 40' beams, the
corresponding A-B amplitudes are 3.6\% and 14\% for 1 arcminute and 4
arcminute offsets. Residuals for beam offsets are plotted in
Figure~\ref{fig:pairoff}.  The effect is less than a few percent for
$\ell < 80$ for the largest offset case.

\subsection{Beam Systematics}

\begin{figure*}
\centering \includegraphics[width=10cm,clip,angle=-90]{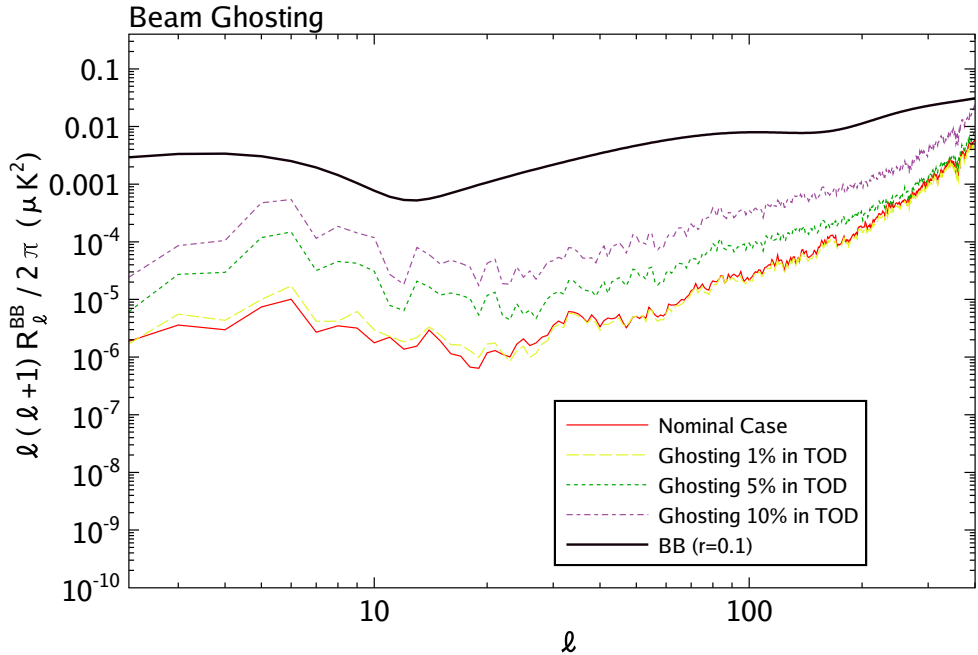}
\caption{\small The residuals from beam ghosting in the \spider\ optics.
The ghosting is added in terms of a percent contamination to the nominal
time-stream.  Final maps are then reconstructed assuming no reflection
contamination with no attempt to correct for the image distortion.
A 10\% contamination in the TOD yields a BB fractional residual as high
as 20\%.  For 5\% ghosting in the TOD this effect is already
down by more than half and is negligible for 1\% contamination in the TOD.
}\label{fig:ghost}
\end{figure*}

The \spider\ antenna array and optics define highly symmetric beams on
the sky.  The beam pattern shown in
Figure~\ref{fig:acbs} is the {\it feed} beam pattern.  The beam on the
sky is influenced by the telescope.  While the
\spider\ telescope edge taper is modest, the beam on the sky will be
more symmetric than the feed pattern shown here.  In particular, the
visibly large and asymmetric lobes above will not propagate to the sky. 
The largest amplitude beam effect
expected comes from reflections or ``ghosting'' in the
\spider\ optics.  Ghosting is common in refractive optics and results
from unintended multiple reflections in the optics.  The effect is a
smaller amplitude beam image which is mirrored with respect to the pixel
position from the centre of the focal plan.

The ghosting is simulated by summing two time-streams from two beams.
One of the time-streams is constructed using offset pointing from the
ghost beam.  The offset pointing of the ghost is determined in
instrument coordinates (elevation and azimuth), hence the $\psi$ angle
for the ghost pointing is calculated appropriately as the final
projection of the orientation of a detector on the sky.  The second
time-stream is constructed from the pointing from the nominal beam.  The
two time-streams are summed with various weighting schemes depending on
the ghost beam contamination\footnote{In order to completely model the
polarization of the ghost, a half-wave plate angle dependency should be
included in the ratio of ghost amplitude to main-beam response.  This is
not done here but is left for future work.}.

The residuals from beam ghosting are summarized in
Figure~\ref{fig:ghost}. For each case the ghosting effect is added in
the time-stream in terms of a percent contamination added to the nominal
time-stream.  Final maps are then reconstructed assuming no reflection
contamination; there is no attempt to correct for the image distortion.
With 10\% contamination in the TOD the effect in the BB residual is as
much as 20\%.  For 5\% ghosting in the TOD this effect is already down
by more than half and is negligible for 1\% in the TOD.  Since
simulations are done only for a single row (of 8 pairs of detectors) the
distance between the original and ghost images is, on average, smaller
for this row than for any other.  It is therefore worth noting that the
full focal plane may show a larger effect than that simulated here.

\subsection{Calibration Drift}\label{sec:drift}

\begin{figure*}
\centering \includegraphics[width=10cm,clip,angle=-90]{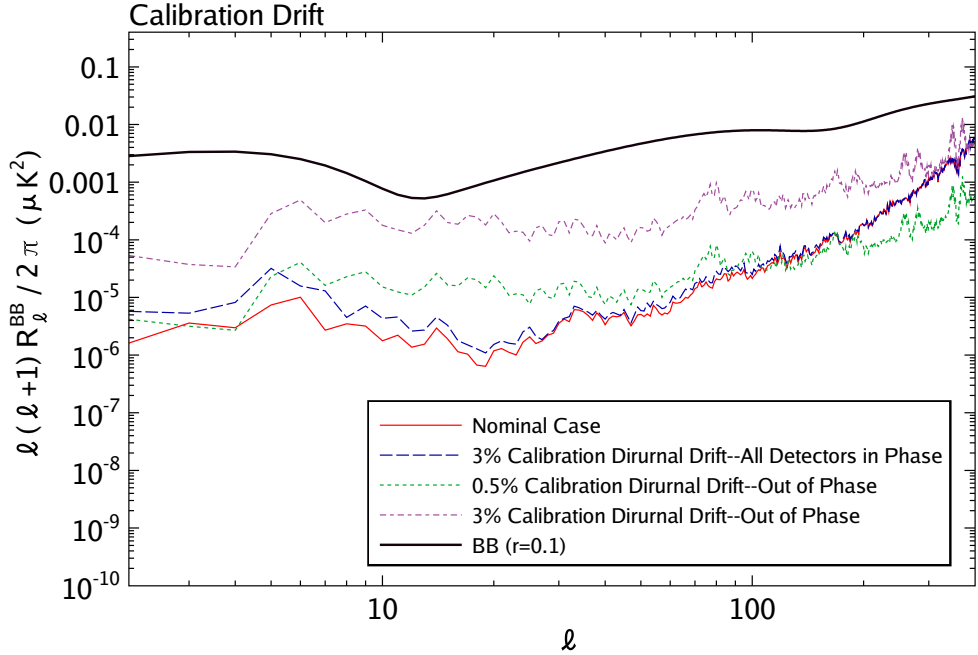}
\caption{\small 
The effect of uncorrected calibration drift.
In the first case the calibration drift is
the same for all detectors, changing on a diurnal timescale with a maximum
amplitude of 3\%.  For the second case the gain drift for all detectors are of the same
amplitude and diurnal but out of phase.  For this case, the BB fractional residual is
as much as 50\% for a gain drift amplitude of 3\% and drops below 5\%
for a 0.5\% amplitude gain drift.  Simulations consider only 8 pairs in a single column or 16 detectors
total.  The RMS result for the full focal plane will average down as
$\sqrt{N}$, where $N$ is the number of detectors.  For the second case
(out of phase gain drifts) a factor of $\sqrt{16}/\sqrt{1024}$ (assuming 1024 CMB science channels) 
has been applied to each residual result. 
}\label{fig:gain}
\end{figure*}

Diurnal variations in the detector
sensitivity will occur due to altitude-induced changes in background loading.
These sensitivity changes will be tracked using 4K
semiconductor emitters (fired intermittently) similar to those
used in \boomerang\ flights. For \boomerang\ 2003 flight, the gain drift
for each individual detector was determined by the uncertainty on an individual
calibration pulse and was measured to be 0.05\%. With an improved version of the
calibration lamp and \spider\rq s higher sensitivity,
the expected uncertainty is $\sim 0.01$\%. Knowledge of the
detector model (determined in pre-flight testing) will also
allow the calculation of the sensitivity for any given operating point.

Figure~\ref{fig:gain} shows the effect of uncorrected calibration drift.
Two cases are considered.  For the first case the calibration drift is
the same for all detectors, changing on a diurnal timescale with a maximum
amplitude of 3\%.  For this case the resulting residual is small; less
than a percent at all scales.

For the second case the gain drift for all detectors are of the same
amplitude and with a 24 hour period, but each of the 16 detectors have a
gain drift with a different phase.  Thus at any given time sample the
calibration factor within a pair of detectors will be different. 
Simulations consider only 8 pairs in a single column or 16 detectors
total.  Again, the RMS result for the full focal plane should average down as
$\sqrt{N}$, where $N$ is the number of detectors.  For this case a
factor of $\sqrt{16}/\sqrt{1024}$ (assuming 1024 CMB science channels) 
has been applied to each residual result.  The BB fractional residual is as much as
50\% for a gain drift amplitude of 3\%.  For a 0.5\% amplitude drift the
residual drops below the 5\% level.  As with all simulations in this
work there has been no attempt to correct for the gain systematic.  For
\boomerang\ 2003, the final relative calibration uncertainty was
0.4\%~\citep{Masi:2006}. \spider\ is expected to achieve an uncertainty
of 0.1\% or less which will be more than adequate to meet science goals.

\section{Conclusions}\label{sec:conclusions}

The results from Section~\ref{sec:results} are summarized in terms of
experimental specifications in Table~\ref{tab:summary}.  While results
are \spider-specific the order of magnitude of various effects can be
translated to other CMB polarization experiments.  The RMS B-mode signal
for $r = 0.1$ is roughly 10000 $\rm nK^2$ ($\sim$1436 $\rm nK^2$ at
$\ell = 8$ and $\sim$7379 $\rm nK^2$ at $\ell = 80$), and scales
linearly with r.  Experimental specifications are set by limiting the
allowed systematic residual level to a factor of $\sim10$ smaller than the 
B-mode signal for $r = 0.01$.

While the simulations were signal-only the impact of large low
frequency detector noise (1/f noise) is reflected in the large scale 
degradation of the B-mode signal for the stepped half-wave plate mode of
operation.  Rapid, continuous half-wave plate modulation mitigates this
effect entirely.  Rapid, continuous gondola rotation also works but only
with iterative map-making which accurately recovers the larger scale
signal. 

It is important to note that the effects studied in
Sections~\ref{sec:modes} and Sections~\ref{sec:noise} (naive versus
iterated maps, spinning more slowly, stepping half-wave plate versus spinning
half-wave plate) will degrade the signal-to-noise achieved on the
bandpowers.  These effects differ from the systematics studied in
Section~\ref{sec:point} to Section~\ref{sec:drift} (pointing
reconstruction errors, polarization angle uncertainty, uncorrected
ghosting, uncorrected gain drifts) which will ultimately bias the final
result.  The requirements on the biasing effects are more difficult
treat than the signal-to-noise issues.
 
The impact of systematics on B-mode polarimeter experiments is also
discussed in~\cite{hhz:2003} and more recently ~\cite{odea:2007}, where
analytical methods are used for calculating the B-mode spectrum bias.
The results are useful for setting experimental ``benchmark parameters''
at the very earliest phases of instrument design. This work goes a step
further by considering the impact of systematics in the
map/time-domain; A necessary step in the evolution of an experiment
which aims to measure the tiny primordial, gravity wave signal.

\begin{table*}
\centering
\begin{tabular}{|c|c|c|}
\hline\hline
\space
{\bf Systematic} & {\bf Experimental Spec.} &
{\bf Comments} \\
\hline\hline
Receiver & &  for 110dps\\
1/f knee & $< 200$ mHz & gondola spin \\
\hline
Receiver & &  for 36dps\\
1/f knee & $< 100$ mHz & gondola spin \\
\hline
Pointing  & & sufficient for\\
Jitter & $< 10'$ & $\ell < 50$ \\
\hline
Absolute&   &  \\
Pol. Angle Offset & $< 0.25$ deg.  &  \\
\hline
Relative&  & \\
Pol. Angle Offset & $< 1$ deg. &  \\
\hline
Knowledge of &   &  sufficient for\\
Beam Centroids & $< 1'$ & $\ell < 30$ \\
\hline
Optical &  &  \% TOD\\
Ghosting & $< 2\%$ & contamination\\
\hline
Calibration &  &  \\
Drift & $< 3.0\%$ & in phase \\
\hline
Calibration &  &  \\
Drift & $< 0.1\%$ & out of phase \\
\hline\hline
\end{tabular}
\caption{\rm Summary of experimental specifications based on simulation results.  
Realistic-amplitude, time-varying systematics are injected in the
simulated time streams.  Maps are reconstructed without any attempt to
correct for the systematic errors.  Experimental specifications are set by limiting the
allowed systematic residual level to a factor of $\sim10$ smaller than the 
B-mode signal for $r = 0.01$.  The nominal operating mode is a 36 dps gondola spin rate, with
the half-wave plate stepping $22.5^{\circ}$ once per day, with 10
iterations (sufficient to recover the residual levels of the
continuously-rotating half-wave plate case) of the map-maker, a Jacobi iterative
solver~\citep{Jones:2007}. 
}\label{tab:summary}
\end{table*}

\acknowledgments
This research used the McKenzie cluster at CITA, funded by the Canada
Foundation for Innovation. Some of the results in this paper have been
derived using the HEALPix package~[\cite{Gorski:2005}] as well as the
FFTW package~[\cite{fftw}].
\acknowledgments

\bibliographystyle{apj}
\bibliography{ms.bib}

\end{document}